\RequirePackage{ifpdf}
\documentclass[cits]{PoS}
\usepackage{amssymb,fontenc,times,mathptmx,graphicx,natbib,url}
\title{The connection between radio and high energy emission in black hole powered systems in the SKA era}

\ShortTitle{Radio-gamma ray connection in the SKA era}

\author{
\speaker{M.\ Giroletti}$^a$, M.\ Orienti$^a$, F.\ D'Ammando$^a,b$, F.\ Massaro$^c$, G.\ Tosti$^{d}$ on behalf of the \fermi-LAT collaboration, 
and R.\ Lico$^{a,b}$, G.\ Giovannini$^{a,b}$, I.\ Agudo$^{e}$, A.\ Alberdi$^{f}$, H. Bignall$^{g}$,
M.\ Pandey-Pommier$^{h}$, A.\ Wolter$^{i}$\\
$^a$INAF Istituto di Radioastronomia, Bologna, Italy;
$^b$University of Bologna, Italy;
$^c$University of Turin, Italy;
$^d$University of Perugia, Italy;
$^e$Joint Institute for VLBI in Europe, The Netherlands;
$^f$Instituto de Astrofisica de Andalucia, Spain;
$^g$Curtin Institute of Radio Astronomy, Australia;
$^h$Observatoire de Lyon, France;
$^i$INAF-Osservatorio Astronomico di Brera, Italy

        E-mail: \email{giroletti@ira.inaf.it}}

\abstract{
Strong evidence exists for a highly significant correlation between the
radio flux density and $E>100$ MeV gamma-ray energy flux in blazars
revealed by the \fermi\ Gamma-ray Space Telescope. However, there are
central issues that need to be clarified in this field: what are the
counterparts of the about 30\% of gamma-ray sources that are as yet
unidentified? Are they just blazars in disguise or they are something more
exotic, possibly associated with dark matter? How would they fit in the
radio-gamma ray connection studied so far?
With their superb sensitivity, SKA1-MID and SKA1-SUR will help to resolve
all of these questions. Even more, while the radio-MeV/GeV connection has
been firmly established, a radio-VHE (Very High Energy, $E>0.1$ TeV)
connection has been entirely elusive so far. The advent of CTA (Cherenkov
Telescope Array) in the next few years and the expected CTA-SKA1 synergy
will offer the chance to explore this connection, even more intriguing as
it involves the opposite ends of the electromagnetic spectrum and the
acceleration of particles up to the highest energies.
We are already preparing to address these questions by exploiting data from
the various SKA pathfinders and precursors. We have obtained 18 cm European
VLBI Network observations of $E>10$ GeV sources, with a detection rate of
83\% (and higher than 50\% for the unidentified sources). Moreover, we are
cross correlating the \fermi\ catalogs with the Murchinson Widefield Array
commissioning survey: when faint gamma-ray sources are considered, pure
positional coincidence is not significant enough for selecting counterparts
and we need an additional physical criterion to pinpoint the right object.
It can be radio spectral index, variability, polarization, or compactness,
needing high angular resolution in SKA1-MID; timing studies can also reveal
pulsars, which are often found from dedicated searches of unidentified
gamma-ray sources. SKA will be the ideal instrument for investigating these
characteristics in conjunction with CTA. 
A proper classification of the unidentified gamma-ray
sources and the study of the radio-gamma ray connection will be essential
to constrain the processes at work in the vicinity of super massive black
holes. 

}

\FullConference{
Advancing Astrophysics with the Square Kilometre Array\\
June 8-13, 2014\\
Giardini Naxos, Italy}

\newcommand{\skipthis}[1]{}

\newcommand\fermi{{\it Fermi}}

\begin{document}

\section{Introduction}

Blazars are the rarest and most extreme class of Active Galactic Nuclei (AGN). They are radio loud sources, characterized by the presence of a relativistic pair of plasma jets whose axis is closely aligned with our line of sight. The ultimate source of their huge energetic output (bolometric isotropic luminosity reaching up to $10^{49}$ erg s$^{-1}$ in the most extreme events) is the gravitational potential of a super massive black hole (SMBH, with mass up to $M_\mathrm{BH}\sim 10^9 M_\odot$). Blazars radiate across the entire electromagnetic spectrum, with a typical two-humped spectral energy distribution (SED). On the basis of their optical spectra, blazars are divided into flat spectrum radio quasars (FSRQs, with prominent emission lines) and BL Lac type objects (BL Lacs, with featureless optical spectra). Moreover, blazars are often classified according to the peak frequency of the synchrotron component of their SED, as low-synchrotron peaked (LSP, with $\nu_\mathrm{peak}\le10^{14}$ Hz), intermediate-synchrotron peaked (ISP, with $10^{14}\ \mathrm{Hz} \le \nu_\mathrm{peak}\le10^{15}$ Hz), or high-synchrotron peaked (HSP, $\nu_\mathrm{peak}\ge10^{15}$ Hz) sources, with FSRQs being in general of the LSP type.  In addition to the different optical spectra and SED, the two subclasses have other differences in observational and evolutionary properties.

Despite being the rarest class of AGN, blazars dominate the census of the gamma-ray sky at high and very high energy (VHE, $E>0.1$ TeV). In the last years, there has been a tremendous progress in the understanding of the physical properties of blazars thanks to the advent of the \fermi-Large Area Telescope (LAT) and its synergy with a large number of multi-wavelength (MWL) facilities across the electromagnetic spectrum. In particular, monitoring projects in the radio (MOJAVE, Boston University monitoring, F-GAMMA, VIPS) have provided strong information about blazar physics, and in some cases possibly imaged the gamma-ray emission zone itself \citep[e.g.][]{Agudo2011}. With further advances expected thanks to the construction and operation of the Cherenkov Telescope Array (CTA) at VHE, significant efforts remain necessary for a conclusive answer on the many open questions in blazar physics, including fundamental issues such as where and how the gamma-ray emission is produced. Moreover, a significant fraction of gamma-ray sources, both at HE and VHE, lacks a proper association to known astrophysical objects; it is important to understand if these sources are just blazars in disguise or more intriguing kinds of objects, possibly related to dark matter annihilation or other exotic physical processes.

The Square Kilometre Array (SKA) will be a fantastic instrument to help address these points. Indeed, understanding the blazar phenomenon inevitably requires a characterization of the physical properties of relativistic jets, which are the subject of other Chapters of this book \citep[e.g.][]{Agudo2014a}. In the present Chapter, in particular, we plan to highlight how the SKA project will shed light on the properties of gamma-ray sources as a population, starting from ongoing and incipient works with pathfinders and precursors, continuing with the SKA itself in its phase 1 and ultimately with the full sensitivity and frequency range that will be available in phase 2. This Chapter is organized as follows: in Section \ref{s.gamma}, we present an outline of the current and planned HE and VHE surveys; in Section \ref{s.results}, we review the state of the art about the connection between radio and gamma-ray emission (\ref{ss.radiogamma}) and the search for unidentified gamma-ray source (UGS) counterparts (\ref{ss.ugs}); a sample ongoing topics are presented in Section \ref{s.ongoing}; in Section \ref{s.ska}, we finally deal with the prospects for SKA in these areas. Further topics are discussed in other Chapters of this book \citep[][]{Agudo2014a,Bignall2014,Corbel2014,Donnarumma2014,Paragi2014,Turriziani2014}, as the connection between radio and high energy emission in black hole powered system (both stellar and super massive) is of great interest and has implications on many other areas of astrophysics.





\section{Gamma-ray instruments and census\label{s.gamma}}

\subsection{High energy gamma rays (10 MeV $<E<$ 300 GeV)}

Blazars have dominated the census of gamma-ray sources (particularly at high galactic latitude) since the Compton Gamma Ray Observatory (CGRO) mission in the 1990's. This evidence has further strengthened in recent years, thanks to the ongoing AGILE \citep{Tavani2009} and \fermi\ \citep{Atwood2009} missions. In particular, the LAT onboard \fermi\ is a pair-conversion gamma-ray telescope sensitive to photon energies from about 20 MeV up to $>300$ GeV, with a large field of view (2.4 sr) and a relatively narrow point spread function (PSF) of $0^\circ.6$ (for a single photon, at 1 GeV). For most of its operations since launch, the LAT has operated in all-sky survey mode, completing a number of catalogs with increasing sensitivity \citep{Abdo2009a,Abdo2010a,Nolan2012} and providing the opportunity to study gamma-ray light curves and time resolved spectra at a variety of time scales for a large number ($\sim10^3$) of sources. The most recent release is the third \fermi\ catalog of gamma-ray sources, based on 4 years of data, and consisting of $\sim 3\,000$ sources \citep{The Fermi-LAT 
Collaboration2015}. 

The AGN population of each \fermi\ catalog is described in accompanying dedicated papers, the most recent of which is the second LAT AGN catalog \citep[2LAC,][]{Ackermann2011b}. In the 2LAC ``clean'' sample\footnote{The ``clean'' sample contains only high-latitude ($|b|>10^\circ$) sources, with a single AGN association, and without analysis flags. The total number of 2FGL sources associated with AGNs is 1121, i.e.\ 60\% of the total.}, there are 886 active galaxies, accounting for 47\% of the total number of \fermi\ sources. Of these 886, 395 are BL Lacs, 310 are FSRQs, 157 are blazars of unknown type (typically because of the lack of an optical spectrum of adequate quality), and 24 are AGNs of other type. The latter group includes 8 so-called misaligned AGNs \citep[typically radio galaxies with small-to-intermediate jet viewing angle, see e.g.][]{Abdo2010b},  4 radio loud narrow-line Seyfert 1 \citep[NLS1,][]{Abdo2009b}, 2 galaxies dominated by starburst activity, and 10 more sources not yet fully characterized.  \fermi\ BL Lacs and FSRQs differ in many observational properties, such as photon index (with the BL Lacs being on average harder than FSRQs), luminosity (with FSRQs being more powerful than BL Lacs), and variability; the most distant \fermi\ FSRQs in the 2LAC is located at a redshift $z=3.10$, while BL Lacs have typically lower redshift, as HSP in particular show strong negative evolution \citep{Ajello2014}; a significant fraction of BL Lacs however lack a redshift measurement. When sources are classified according to the peak frequency of the synchrotron component of their SED, the numbers of LSP, ISP, and HSP blazars in the 2LAC are 301, 95, and 213, respectively.

Beside blazars, pulsars are the second largest population of associated gamma-ray sources in the 2FGL; interestingly, pulsars are also generally bright radio sources and they are indeed one major subject for SKA studies, as discussed in other Chapters. A few systems associated to accretion and ejection around stellar mass black holes (including microquasars, X-ray and gamma-ray binaries) are also found at high energy. However, the total number of pulsars and other associated sources besides blazars reaches only 195 (55 at $|b|>10^\circ$), while the majority of the remaining sources are so far unassociated (31\% of the total). The existence of a large fraction of UGSs has been around since the EGRET era, although its composition has changed a lot given the higher sensitivity and better localization provided by \fermi, and the vastly improved MWL catalogs at all other wavelength.

\subsection{Very high energy gamma rays ($E>100$ GeV)}

At larger energies, the sensitivity of space detectors is not adequate and observations are carried out on the ground through the detection of Cherenkov radiation triggered by VHE photon interactions in the atmosphere. Current generation of VHE Imaging Atmospheric Cherenkov Telescopes (IACT) exploits the stereoscopic technique to improve sensitivity and characterization of the event, and is implemented in three major observing arrays MAGIC, HESS, and VERITAS. These arrays and their predecessors have so far detected 147 sources, which are listed in the so-called TeVCat catalog\footnote{\url{http://tevcat.uchicago.edu/}}.

The census of VHE sources is again dominated by (occasionally misaligned) blazar sources, with BL Lacs, FSRQs, and FRI radio galaxies accounting for 38\% (58/147) of the total population (a fraction that becomes significantly larger at high galactic latitudes). However, the overall properties of the VHE blazar population are markedly different from those at lower energy. BL Lacs, and in particular HSP blazars, are the most numerous class, with 50 detected sources, 42 of which are of the HSP type. Only 3 FSRQs are known, with the most distant one being 3C\,279 at $z=0.536$.
The typical luminosity is much lower than for \fermi\ blazars, and the redshift distribution is also very different, being most sources at a very low redshift. This is both a consequence of their physics and of the absorption through photon-photon interaction with the extragalactic background light (EBL) being more pronounced for more distant sources.

The fraction of unidentified sources is 18\%. One significant difference between  IACTs and \fermi\ is that the latter operates in survey mode, while the formers generally do pointed observations. This generally produces a bias in the selection of the VHE targets, and it is to be expected that many more unidentified sources would be discovered in a blind survey, which could be possible with the advent of CTA.

\section{State of the art\label{s.results}}

\subsection{The radio-gamma ray connection}\label{ss.radiogamma}

\begin{figure} 
\begin{center} \includegraphics[width=.55\textwidth]{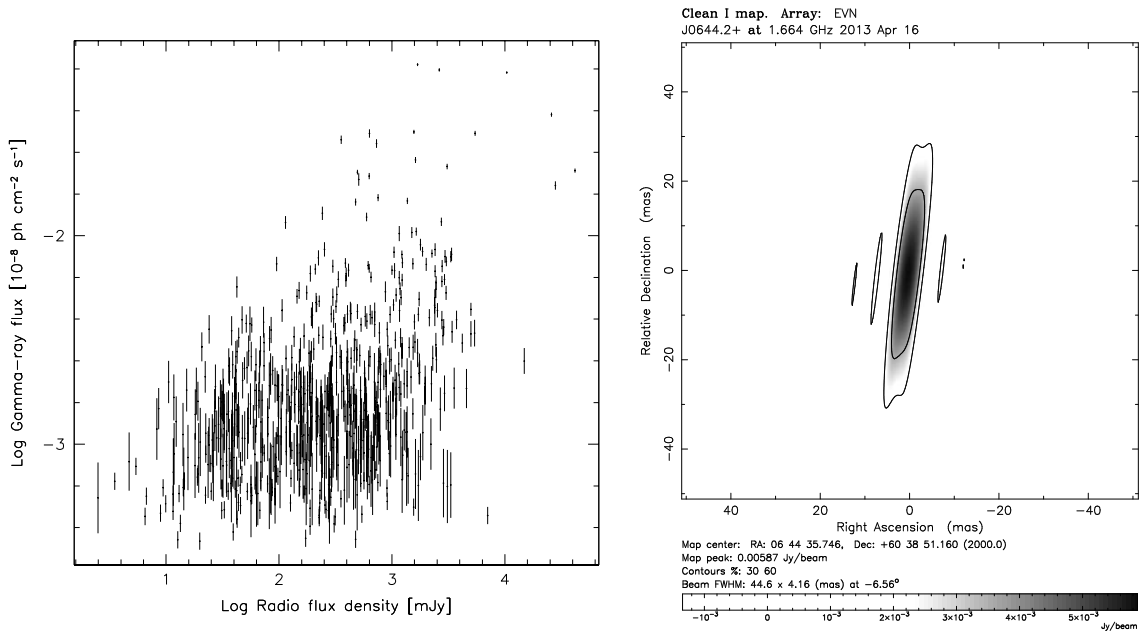}
\includegraphics[width=.44\textwidth]{j0644.eps} \caption{Left: gamma-ray vs radio flux for 1LAC sources, adapted from \citet{Ackermann2011a}. Right: EVN 18cm image of NVSS J064435+603849, showing a compact source possibly associated to the UGS 2FGL J0644.2+6036.} \label{figura} 
\end{center}
\end{figure}

As radio loud sources dominate the extragalactic gamma-ray population, it is clear that there is a connection between radio and high energy emission, physically originated by the presence of a relativistically beamed jet emerging from a SMBH. However, not all blazars are detected in gamma rays, and in particular at VHE the most luminous radio sources are hardly, if ever, detected by current instruments. 
Several works since the 1990's have attempted to reveal the details of such a connection \citep[e.g.,][]{Padovani1993,Stecker1996,Muecke1997,Ghirlanda2010,Mahony2010}. The most extensive work dealing with the radio/gamma-ray connection at MeV/GeV energy is the one by \citet[][see also Fig.~\ref{figura}, left panel]{Ackermann2011a}, based on 599 sources detected in the first catalog of \fermi\ AGN \citep[1LAC,][]{Abdo2010c}. In this work, we applied a dedicated statistical tool \citep{Pavlidou2012} to establish the strength and significance of the correlation between radio flux density and gamma-ray energy flux using archival and concurrent radio data, considering different gamma-ray energy bands, and analysing the whole \fermi\ AGN population as well as all the different blazar sub-classes. The main findings are as follows: radio flux density and gamma-ray energy flux are correlated with a very high significance (chance probability $<10^{-7}$) in the whole 1LAC blazar sample, as well as for BL Lacs and FSRQs separately; gamma-ray data correlate better with concurrent radio data than with archival ones; there is a low significance hint of a trend between blazar SED type and strength of the correlation in gamma-ray sub-bands.

However, even in the most significant cases, the correlation itself shows a large scatter, and the distribution of data points is very broad, especially at low fluxes and near the sensitivity limit of the \fermi\ survey. Therefore, it is misleading to use such correlation to extrapolate gamma-ray luminosity functions from the radio ones. Future improvements could be obtained through the use of substantially deeper gamma-ray data (such as those that will be available at the end of the \fermi\ mission) and/or shorter wavelength radio emission, which is produced on regions closer to the gamma ray emission site and could provide a tighter correlation \citep{Agudo2014b}. In any case, direct use of the blazar gamma-ray luminosity function will remain more accurate, while estimate based on the radio-gamma correlation might apply to radio galaxies and star forming galaxies.

As far as the radio-VHE connection is concerned, the situation is even more complex. Physically, it is natural to expect an anti-correlation between VHE and radio luminosity, since the most powerful radio sources tend to have low-frequency peaked SEDs, soft gamma-ray photon indexes, and to suffer strong EBL absorption due to their large distance. Moreover, the IACTs operational mode introduces significant biases in the target selection, so that a complete survey and a statistically significant assessment of any radio-VHE correlation are precluded. In any case, it is still confirmed that the majority of VHE detected sources are indeed radio loud blazars.

\subsection{Un-associated gamma-ray sources\label{ss.ugs}}

Overall, about 30\% of \fermi\ sources lack a high-confidence low-frequency counterpart. This fraction is not constant among gamma-ray flux bins, as it becomes larger for faint gamma-ray sources. First, the gamma-ray error ellipse is larger; moreover, on the basis of the radio-gamma ray correlation described above, faint gamma-ray sources are likely associated with low flux density radio sources, whose space density is larger and which often lack an optical spectrum. Therefore, it is likely that many weak gamma-ray blazars remain unrecognized because their low-frequency counterpart can not reach the statistical significance necessary to call a formal association. There are various ways to get around this difficulty at present. The logical approach would favour the use of all-sky high-frequency radio surveys, as radio sources other than blazars are rare at high frequency. However, large and deep high-frequency surveys are very difficult in practice: the most valuable resource so far in this field is the Australia Telescope 20 GHz (AT20G) survey, with its catalog going down however only to a flux-density limit of 40 mJy \citep{Murphy2010}.

A complementary approach aims to the selection of blazars based on some characteristic spectral features, such as the infrared colors or the low frequency spectral index. In this way, it is possible to exploit existing large and comparatively deep surveys, where blazars are generally a minority population. For example, \citet{Massaro2011} and \citet{D'Abrusco2013} have shown how blazars form a so-called {\it strip} in the IR color space and have exploited this feature to propose new IR counterparts for $\sim 20\%$ of the UGSs investigated \citep{Massaro2013a}. Similarly, \citet{Massaro2013b,Massaro2013c} have demonstrated that blazars maintain a flat spectral index well below $\nu<1$ GHz, and proposed 31 new blazar associations \citep[see also][]{Nori2014}. We are also investigating the low frequency morphology of \fermi\ blazars through high angular resolution images obtained at the Giant Metrewave Radio Telescope (GMRT).

\section{Some pathfinder projects \label{s.ongoing}}

\subsection{EVN survey of 1FHL sources}

Before dealing with the SKA outcomes in this area, it is worthwhile to highlight some of the results that are being made possible by the SKA pathfinder and precursors. In section \ref{ss.radiogamma}, we briefly discussed how the assessment of a correlation between radio and VHE emission is affected by the IACT observational mode. Eventually, CTA will offer the opportunity to overcome this difficulty thanks to a uniform survey simultaneous to the SKA operations; at present time, the most useful resource in this field is probably the first \fermi-LAT catalog of 514 sources detected above 10 GeV \citep[1FHL,][]{Ackermann2013}, based on LAT data accumulated during the first three years of survey. While of course not being entirely representative of the VHE population, the 1FHL indeed provides a sample whose average properties are quite different from those of the full-band \fermi\ catalogs: the SED-based classification starts to lean towards HSP types (with 99, 71, and 162 sources of the LSP, ISP, and HSP types respectively); the redshift distribution only extends out to $z=2.5$, and the overall luminosities are generally lower. The fraction of unidentified sources is low, just 13\%. 

We are now investigating the 284 1FHL sources at $\delta>0^\circ$ with the use of high angular resolution Very Long Baseline Interferometry (VLBI) radio observations. We note that VHE blazars tend to display peculiar VLBI properties, e.g.\ comparatively low brightness temperature and lack of superluminal motions \citep{Piner2008,Lico2012}, which are otherwise common in MeV/GeV blazars. In particular, we have observed with the European VLBI Network (EVN) at 1.7 GHz the sky regions of the 71 northern 1FHL sources without any existing VLBI data but with at least one NVSS source in the \fermi\ error ellipse. Of these, 49 are classified as blazars, one is a supernova remnant, the remaining 21 are unassociated sources. While a detailed presentation of the observations and discussion of the results is given in a dedicated paper \citep{Giroletti2014}, we highlight here some of the most relevant finding, also in perspective of actual SKA observations. For sources classified as blazars, we have a detection rate of 100\%, confirming that parsec scale cores are ubiquitous in gamma-ray blazars; eventually, this will allow us to explore the existence of a correlation between radio and $E>10$ GeV gamma-ray flux using high resolution, concurrent data, which is clearly desirable. For the unassociated sources, we have a relatively high detection rate (preliminarily, 11/21) of compact components; given the low space density of milliarcsecond scale sources, they are most likely the true counterparts of the 1FHL sources, and our data provides important information for follow-up studies and a proper classification. Interestingly, among the unassociated sources there are four objects that are also present in the 2FGL and for which there is a blazar candidate selected with the low-frequency spectral index method described in \ref{ss.ugs}:\ for all of them, we find a compact radio source, which further supports their blazar nature and the validity of the method (see e.g.\ NVSS J064435+603849 in the right panel of Fig.~\ref{figura}).

\subsection{MWA-\fermi\ catalog cross correlation}

Blazar catalogs become rather incomplete around $\sim 10$'s mJy flux density levels, because of the technical difficulty in going deep at high radio frequency and of the large number of contaminating steep-spectrum radio sources at low radio frequency. Additional constraints can be used to narrow the number of blazar candidates at low frequency, like we did on the basis of the flat radio spectrum. Thanks to the SKA pathfinders and precursors, there will soon be several sensitive, wide area, low frequency surveys, ideal for this scope. In particular, even in its early commissioning phase, the Murchinson Widefield Array (MWA) has surveyed approximately 9 hr in R.A.\ over a declination range of $45^\circ$ at three different frequencies (119, 150 and 180 MHz). Of order 10,000 sources were detected at $5\sigma$ above the confusion noise limit, determining a flux and spectral index for each of them \citep{Hurley-Walker2014,Morgan2014}. We are currently cross-correlating this catalog against the 2FGL and 3FGL, with the goals of (1) verifying whether the radio-gamma ray correlation extends at very low radio frequency, (2) characterizing the spectral properties of known blazars in this unexplored window of the electromagnetic spectrum, and (3) searching for counterparts to unidentified gamma-ray sources among the low frequency population.

\section{SKA prospects \label{s.ska}}

SKA will surely be a game changer for the topics described in this Chapter,
starting from its Phase 1 (SKA1) and even in the early science phase. We can anticipate at least two major well-defined projects that shall be carried out in phase 1: the first one (\ref{ss.ska1a}) deals with the identification and characterization of as large a fraction as possible of the MeV/GeV and the VHE populations, and the second one (\ref{ss.ska1b}) does then address the implications on the physics of gamma-ray sources, and in particular of relativistic jets supported by SMBH.

For SKA1 specifications, we consider SKA1-MID observations in Bands 1, 3, and 5 (350-1050 MHz, 1.65-3.05 GHz, and 4.6-13.8 GHz), and SKA1-SUR data from Phased Array Feed (PAF) in the 1.5-4.0 GHz range. Since blazars are continuum emission sources with flat spectrum, the choice of bands can reasonably be adjusted; however, it would be highly desirable to keep the highest frequency bands (4.6-13.8 GHz for SKA1-MID, and possibly 1.5-4.0 GHz for SKA1-SUR), where there is less contamination from steep spectrum sources and de-polarization effects are less severe. Indeed, polarization has not been discussed so far in this Chapter but it will be a very important probe to the physics (see also the Chapter by Agudo et al.). The full band and the longest baselines provided by SKA2 will provide further critical information (\ref{ss.ska2}).

\subsection{SKA1 \& the identification and characterization of gamma-ray sources \label{ss.ska1a}}

The fraction of UGS in the MeV/GeV domain is around 30\% at present and it is essential to understand how many of these sources are known gamma-ray emitters yet to be identified and how many are related to unknown classes and/or to exotic physics. By the time SKA will be operational, \fermi\ shall have completed at least a 10 year survey, revealing $\sim 10^3$ UGS. At the same time, CTA will be starting its operation, and in particular if it will carry out even a moderately shallow wide area survey, it will reveal a large number (probably several $\times10^2$) of new, possibly weak and/or transients VHE sources. While continuum sensitivity of existing interferometers is already reasonably suited for the detection of classical gamma-ray blazars, the SKA1 characteristics (wide field, polarimetry, bandwidth, timing) will be essential for a full characterization of the UGS sources. 

A first, fundamental step will consist in the build up of a complete and deep blazar catalog, {\it  based on SKA1-SUR in the band 1.5-4.0 GHz}. Sensible choices for selecting blazars in this frequency range would be a spectral index $\alpha<0.5$ and a polarization fraction $p>0.5\%$. Assuming an instantaneous bandwidth of 500 MHz and the typical PAF sensitivity, it will be possible to extend surveys such as the Combined Radio All-Sky Targeted Eight GHz Survey \citep[CRATES,][]{Healey2007} by at least one magnitude, going down to $\sim 5$ mJy completeness, with the additional benefit of polarization information. CRATES has been a powerful resource in identifying blazars in the early releases of the \fermi\ catalogs but it has become more and more incomplete as counterparts became fainter and fainter. In addition to pulsar searches which will further diminish the number of UGS, the fraction of UGS will be significantly reduced. 
\citet{Ackermann2012} used a statistical method to propose a blazar or pulsar origin for more than half of the 1FGL UGS, and MWL data generally confirm the proposed classifications. Similarly, we can predict that a significant fraction (possibly around 50\%) of the still unassociated \fermi\ sources will be actually classified thanks to SKA1-SUR observations. We also expect a strong synergy with total intensity and polarization surveys from the radio continuum and cosmic magnetism working groups. Identification of transient sources such as microquasars or novae star will also be facilitated by SKA1, already in its early operation phase.

The build up of a complete gamma-ray blazar (and pulsar) catalog will have important scientific implications by itself (see \ref{ss.ska1b}); moreover, it will allow us to select by difference a population of truly exotic gamma-ray sources, belonging to so far unknown classes of gamma-ray emitters. For both the newly discovered blazars and the remaining ``exotic'' UGS, dedicated {\it multi-$\lambda$, multi-epoch SKA1-MID observations} will provide a better characterization of their spectral, variability, structural, and polarization properties. The same approach will of course be adopted for VHE sources detected by CTA, with the additional benefit of simultaneous operations between the two instruments: SKA and CTA will have significant analogies, from the large collecting areas, the remarkable jump in sensitivity with respect to existing instrumentation, the capability to detect transient sources, the multi-fold design (with different parts of the instrument working with different specifications and science goals), the requirements in terms of computational and data storage resources, and the international nature of the projects themselves.

\subsection{SKA1, the radio-gamma ray connection and its physical implications \label{ss.ska1b}}

As a result of a longer exposure as well as a better characterization of both the instrument response function and the diffuse background model \citep{Grove2014}, the final 10-yr catalog of \fermi\ blazars will likely reach a sensitivity limit nearly one magnitude deeper than the 1LAC. An even larger improvement is expected in the VHE band thanks to CTA. With high quality radio data from SKA1 it will thus be possible to clarify the still open points on the radio-gamma ray correlation in the MeV/GeV band and to address for the first time the one with VHE gamma rays. There are various issues for which a definite answer could be reached: the different trends obtained for blazar spectral types, redshift bins, gamma-ray energy bands, and epoch of multi-$\lambda$ data will in turn provide constraints on the processes of high energy emission, blazar evolution, size of the emitting region, duty cycle and variability nature of blazars. For instance, the comparison of polarized radio flux density and gamma-ray energy flux can reveal if and how the magnetic field intensity is actually relevant for the gamma-ray emission (through inverse Compton scattering on synchrotron photons), a link that can not be studied systematically with the sensitivity of present instruments. Moreover, and in particular for the radio-VHE connection, simultaneous radio and gamma-ray observations should be relatively easy with an SKA-CTA synergy thanks to the sensitivity, flexibility, and sub-arraying capability of SKA1-MID; this will reveal how much of the scatter in the correlation is due to non-simultaneity, and therefore provide insights on the size and relative distance between the radio and gamma-ray emission regions. Additional statistical tests could be carried out by comparing the gamma-ray properties not only to the radio flux density but also to other properties such as the spectral index and the polarization percentage, which can trace the compactness, core dominance, and magnetic field configuration of the gamma-ray emitting zone.

For all the above goals it is critical to obtain radio data of high quality for a large number of sources: sub-mJy, polarization sensitive, arcsecond resolution, multi-frequency and possibly multi-epoch data for $\sim 2\,000$ sources (we focus only on the southern sky) will be necessary. It is important to note that the new sources with a low flux density could be either high redshift or low power sources. Either way, they would not just provide an increased statistic for existing studies but actually probe new regions of the space of parameters with respect to currently known gamma-ray blazars. A full survey of this population would be well feasible with SKA1-MID: the faintest blazar in the 2LAC (2FGL J0912.5+2758) has a flux density $S=4.2$ mJy; we can conservatively assume a value about $10\times$ lower for the faintest blazar detected by \fermi\ in its entire mission. Based on a 30 $\mu$Jy s$^{-1}$ minimum detectable flux density for SKA1-MID Band\,5  \citep[][Table 9]{Dewdney2013}, the time requested for a survey of all the visible \fermi\ AGNs down to the faintest one would be mostly driven by slewing time. With clever scheduling and sub-arraying, no more than a few hours would be necessary. If we aim for detection of polarized emission at the 1\% level, only the faintest $\sim100$ or so of blazars would require non-negligible observing time, but the entire project would not be significantly expanded. Even in an early science phase providing $\sim50\%$ of the SKA1 baseline design, this project could be carried out without significant trouble; actually, this is an ideal project for early SKA1 phases since it could permit (nearly) concurrent SKA1-\fermi\ observations.

Finally, we note that the radio-gamma ray scatter plot has a significant spread based on current data; therefore, it can not be used to constrain the blazar contribution to the extragalactic gamma-ray background based on radio luminosity function. However, with deeper surveys and better characterization of the variability in radio and gamma rays, it could become possible to improve our understanding of such connection \citep{Bignall2014}. Eventually, this would allow us also to constrain the available fraction for other more or less exotic contributors (from radio galaxies to dark matter).

\subsection{Full SKA prospects\label{ss.ska2}}

As suggested by variability time scales and SED model fit, the gamma-ray
emission regions are typically very compact (sub-pc) and heavily
self-absorbed at cm-$\lambda$. Therefore, it is critical to obtain high
resolution and high frequency radio data as much as technically feasible. A
direct approach of the gamma-ray region is beyond the scope of SKA and it
requires (sub-)mm-VLBI. Nonetheless, the full SKA will provide the
opportunity to get much closer to this region than SKA1 could. In this
case, the critical requirements will be the completion of the longest
SKA-MID baselines with $\nu>10$ GHz, and possibly an integration with the
African VLBI Network (AVN) and the EVN \citep{Paragi2014}. In particular,
for very high redshift sources ($z\sim5-8$), Band5 observations would
correspond to millimetre wavelength emission in the source's rest frame,
hence providing a tool to observe the innermost regions of blazars.


High resolution, time resolved, sensitive, polarimetric observations of individual sources will become possible, similar to what is currently being carried out only for a handful of special sources, like M87 \citep{Giroletti2012} and 3C120 \citep{Agudo2012}.

\end{document}